\documentclass[12pt]{article}

\usepackage{epsf}
\usepackage[dvips]{graphicx}

\def\bk{{\mbox{\boldmath$k$}}}

\def\bp{{\mbox{\boldmath$p$}}}
\def\bgam{{\mbox{\boldmath$\gamma$}}}

\begin{document}

\begin{center}
{\bfseries PECULIARITIES IN THE STRUCTURE OF TWO-PARTICLE STATES
WITHIN THE BETHE-SALPETER APPROACH}

\vskip 5mm

S.M. Dorkin$^{1}$, S.S. Semikh$^{2 \dag}$, M. Beyer$^{3}$, L.P.
Kaptari$^{2}$

\vskip 5mm

{\small (1) {\it SINP, Moscow State University, Dubna, Russia}
\\
(2) {\it BLTP, Joint Institute for Nuclear Research, Dubna,
Russia}
\\
(3) {\it University of Rostock, Germany }
\\
$\dag$ {\it E-mail: semikh@theor.jinr.ru }}
\end{center}

\vskip 10mm

\section{Introduction}
The two-fermion bound system is an attractive subject of atomic
(positronium, hydrogen atom) and sub-atomic physics (deuteron,
mesons). Despite these systems are rather simple the study of
two-particle bound states is challenging and 
still remains a source of progress in
quantum theory. Last decade significant efforts were undertaken to
describe such objects, and progress has been achieved in both, solving 
the corresponding
equations \cite{gross}-\cite{bakker} and  calculating the
experimentally measured observables \cite{gross1}-\cite{our_brkp}.

The homogeneous Bethe-Salpeter (BS) equation nowadays remains a
powerful tool to investigate the relativistic bound state
problem. Recently the problem of rigorous treating the BS
equations received renewed interest, and several successful
attempts were made to reconsider solving algorithms  for the BS
equation \cite{fb-tjon} and to make its reduction
to Light Front form more transparent \cite{tobias}. A good example of a consistent
study of this subject can be found in Ref. \cite{karm2}.

In fact, our understanding of the mathematical properties of bound
states within any relativistic approach is far from perfection.
The BS equation itself is a quite complicated mathematical object,
and the technical problem of solving it is still of principal
value. In the present paper we propose an efficient and promising
method to solve the BS equation for fermions involving
interaction kernels in the form of one boson exchange supplemented
with the corresponding form factors. It is based on employing the
basis of hyperspherical harmonics for expanding the partial
amplitudes and kernels. We show that this new technique allows
one to utilize many advantages in understanding the BS approach. Basically, the
current study is inspired by the results reported in
\cite{karmanov}. We explore the structure of $0^+$ and $1^+$ bound
states for different couplings studying in details the convergence
of solutions and corresponding eigenvalues. In particular, on a
basis of the introduced rigorous method to solve the BS equation it
becomes possible to analyze in details the problem of stability of
bound states within the BS approach. It is worth mentioning that this
phenomenon was also considered in \cite{karmanov} in the framework
of Light-Front Dynamics, and it is completely similar to the bound
state collapse in non relativistic quantum mechanics for
potentials behaving like $\sim 1/r^2$.

\section{Homogeneous Bethe-Salpeter equation for fermions}
In the present study, we generalize the method described in
\cite{fb-tjon,arhiv,last_yaf} to homogeneous BS equations for
spinor particles. To provide a clear illustration, we present here
the results of solving the equation for the vertex function
\begin{eqnarray}
\label{sphom} {\cal G}
(p)=i\int\frac{d^4k}{(2\pi)^4}\,V(p,k)\,\Gamma(1)\, S(k_1)\,{\cal
G}(k)\,{\tilde S(k_2)}\,\tilde\Gamma(2)
\end{eqnarray}
in the $^1S_0$ channel with
$V(p,k)=g^2/[(p-k)^2-\mu^2+i\varepsilon]$ as one meson exchange
interaction kernel for scalar, pseudoscalar and vector meson
exchanges with corresponding vertices $\Gamma(i), i=1,2$. The
meaning of the introduced quantities $p,k,k_1,k_2$ is the
following: $k_{1,2}=P/2\pm k$, $k=(k_0,\bk), p=(p_0,\bp)$ are the
relative 4-momenta, and $P=(M,{\bf 0})$ is the total 4-momentum of
the bound state in its c.m.s. Spinor propagators of constituent
particles with equal masses $m$ are
\begin{eqnarray}
\nonumber {S}(k)=\frac{\hat k +m}{k^2-m^2+i\varepsilon},\quad
{\tilde S}(k)\equiv C{S}(k)^TC=\frac{\hat k
-m}{k^2-m^2+i\varepsilon},
\end{eqnarray}
with $C$ standing for charge conjugation matrix,
$C=i\gamma_0\gamma_2$, $\hat k=\gamma_\mu k^\mu$. In general, the vertices include the
meson-nucleon form factor, which will be further considered as
\begin{eqnarray}
F(q^2)=\frac {\Lambda^2-\mu^2}{\Lambda^2-q^2}.\label{ff}
\end{eqnarray}
In (\ref{sphom}) the BS vertex function ${\cal G}(p)$ being the
matrix $4\times 4$ should be expanded over the proper set of
matrices for the given channel.  The   standard choice for them is
the set of the $\rho$-spin angular momentum vector harmonics
$\Gamma_\alpha(\bp)$, where index  $\alpha$ includes not only
$LSJ$ momenta but also $\rho$--spin quantum numbers, which are
denoted by $++$,$--$, $e$ and $o$ \cite{kubis}. For convenience in
our further considerations we introduce another equivalent set of
spin-angular matrices instead of $\rho$-spin basis. For $^1S_0$
channel the following functions may be chosen:
\begin{eqnarray}
{\cal T}_1(\bp)=\frac 12 \gamma_5,\quad{\cal T}_2(\bp)=\frac 12
\gamma_0\gamma_5, \quad{\cal
T}_3(\bp)=-\frac{(\bp,\bgam)}{2|\bp|}\gamma_0\gamma_5, \quad{\cal
T}_4(\bp)=-\frac{(\bp,\bgam)}{2|\bp|}\gamma_5.\label{nharms}
\end{eqnarray}
This basis is orthonormal, i.e.
\begin{eqnarray}
\int d\Omega_p\, Tr\,[{\cal T}_m(\bp){\cal
T}_n^+(\bp)]=\delta_{mn} \nonumber,
\end{eqnarray}
and the partial expansion can be written as
\begin{eqnarray}
{\cal G}(p_0,\bp)=\sum\limits_{n}g_n(p_0,|\bp|) \,{\cal
T}_n(\bp),\quad g_n(p_0,|\bp|)= \int d\Omega_p\, Tr\,[{\cal
G}(p_0,\bp){\cal T}_n^+(\bp)]. \label{exp2}
\end{eqnarray}
After the partial expansion and upon performing the Wick rotation
we can expand the partial vertex functions $g_n$ and the partial
interaction kernels in hyperspherical functions using the
formula \cite{fb-tjon}
\begin{eqnarray}
V_E(p,k)&=&-\frac {1}{(p-k)_E^2+\mu^2}=-2\pi^2 \sum_{nlm}
\frac{1}{n+1} V_n(\tilde p,\tilde k)Z_{nlm}(\omega_p)
Z_{nlm}^{*}(\omega_k),\label{exp12}\\ V_n(a,b)&=& \frac
{4}{(\Lambda_+ +\Lambda_-)^2} \left( \frac {\Lambda_+
-\Lambda_-}{\Lambda_+ +\Lambda_-}\right)^n,\quad
\Lambda_{\pm}=\sqrt{(a\pm b)^2+\mu^2},\nonumber
\end{eqnarray}
where $(k,p)_E\equiv k_4\,p_4+(\bk,\bp)$, $\tilde
k=\sqrt{k_4^2+\bk^2}$ is the 4-dimensional absolute value, and
$\omega_k=(\chi,\theta,\phi)$ are the angles of vector
$k=(k_4,\bk)$ in 4-dimensional Euclidean space. The hyperspherical
harmonics are $Z_{nlm}(\chi,\theta,\phi)=X_{nl}(\chi)
Y_{lm}(\theta,\phi)$ (for details see e.g. \cite{last_yaf}).
Corresponding hyperspherical expansions for $^1S_0$ amplitudes are
given by
\begin{eqnarray}
g_{1,2}(ip_4,|\bp|)&=&\sum_{j=1}^\infty g_{1,2}^j(\tilde
p)\,X_{2j-2,0}(\chi_p),\quad g_{3}(ip_4,|\bp|)=\sum_{j=1}^\infty
g_{3}^j(\tilde p)\,X_{2j-1,1}(\chi_p),\label{s0p1}\\
g_{4}(ip_4,|\bp|)&=&\sum_{j=1}^\infty g_{4}^j(\tilde
p)\,X_{2j,1}(\chi_p).\label{s0p2}
\end{eqnarray}
By using these decompositions one can obtain the final system of
1-dimensional integral equations for the coefficient functions
$g_\beta^j(\tilde p)$. This system will be explicitly shown and
discussed in details separately \cite{our_planned}. What is
important now, is that its numerical analysis does not require
large computer resources. This set can be easily transformed to
the system of linear equations. For this aim, firstly, the
infinite summation over hyperspherical components should be
limited to some finite value $N_{max}$. Secondly, to calculate the
integrals a reliable integration scheme is required. Applying
Gaussian quadrature formula, one can get the system of linear
equations with the sought functions defined in the mesh points
\cite{fb-tjon},
\begin{eqnarray}
X=\lambda\, AX,\label{syst}
\end{eqnarray}
where $\lambda=g^2$, and the column
\begin{eqnarray}\nonumber
X^T=([\{g_1^j({\tilde p}_i)\}_{i=1}^{N_G}]_{j=1}^{N_{max}},\ldots,
[\{g_{N_c}^j({\tilde p}_i)\}_{i=1}^{N_G}]_{j=1}^{N_{max}})
\end{eqnarray}
represents the sought solution in the form of a group of sets of
partial wave components $g_\alpha^j, \alpha=1,...,N_c;
j=1,...,N_{max}$ specified on the integration mesh of order $N_G$.
For $0^+$ ($1^+$) state we have $N_c=4$ ($N_c=8$). The matrix $A$
is obtained as a product of partial kernels, the Jakobian of the
transition to the new variables, the weights of the Gaussian mesh
etc. The dimension of $A$ is $N\times N$, where $N=N_c N_G
N_{max}$.

\begin{table}[t]
\[
\begin{array}{|c|c|c|c|c|c|}
\hline \alpha & M & P_{++}(LFD) & P_{--} & P_{o}(LFD) & P_{e}
\\
\hline\hline
1.194 & 1.937 & 1.012 & -1.18\cdot 10^{-3} & -6.63\cdot 10^{-3} & -4.37\cdot 10^{-3}\\[1ex]
1.592 & 1.892 & 1.020 & -2.99\cdot 10^{-3} & -1.07\cdot 10^{-2} & -6.92\cdot 10^{-3}\\[1ex]
1.989 & 1.842 & 1.030 & -6.22\cdot 10^{-3} & -1.46\cdot 10^{-2} & -9.41\cdot 10^{-3}\\[1ex]
2.149 & 1.820 & 1.034 & -8.11\cdot 10^{-3} & -1.61\cdot 10^{-2} & -1.03\cdot 10^{-2}\\[1ex]
2.308 & 1.798 & 1.039 & -1.05\cdot 10^{-2} & -1.75\cdot 10^{-2} & -1.12\cdot 10^{-2}\\[1ex]
2.348 & 1.788 & 1.041 & -1.25\cdot 10^{-2} & -1.80\cdot 10^{-2} & -1.16\cdot 10^{-2}\\[1ex]
2.352 & 1.5 & 1.210 & -0.19            & -1.24\cdot 10^{-2} & -7.77\cdot 10^{-3}\\[1ex]
 \hline
\end{array}
\]\caption{Pseudo-probabilities of partial components in the state
with given $M$, i.e. their contributions to normalization
condition.}
\end{table}

\section{Results}
\begin{figure}[t]
\vskip -1cm \hskip -0.7cm
\includegraphics[height=8cm]{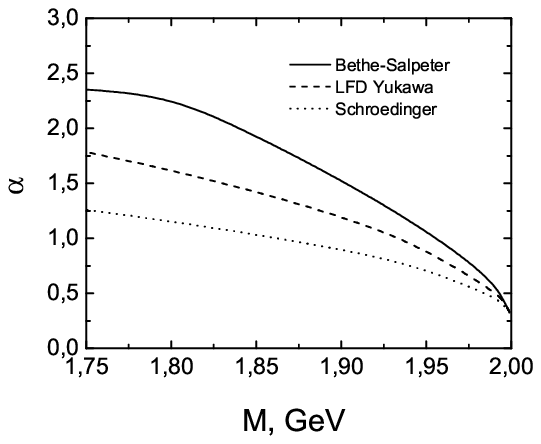}
\hskip -1cm
\includegraphics[height=8cm]{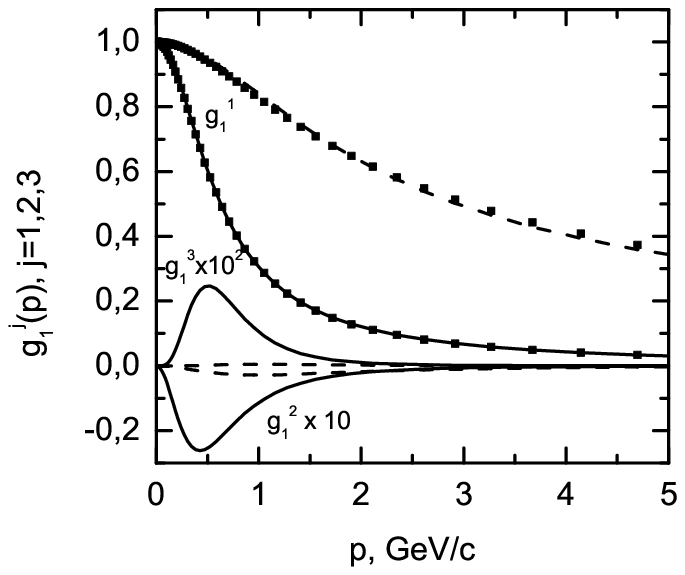}
\\
\parbox[c]{7.5cm}{\raggedright\protect\caption{Dependence of
the coupling constant $\alpha=g^2/4\pi$ from the mass of the bound
state $M$ in different approaches.}\label{pict1}} \hfill
\parbox[c]{7.5cm}{\raggedright\protect\caption{
Functions $g_1^j, j=1,2,3$ for the cut-off $\Lambda=500$ GeV/c and
$g^2=30$ (solid line) and $g^2=48$ (dashed line).}\label{pict2}}
\end{figure}

In this short communication we are able to present only the most
indicative results of the numerical treatment of the BS equations
within our method. First of all, the set of equations (\ref{syst})
has the solution only if $\det(\lambda A-1)=0$. This condition
allows us to connect the coupling constant $g^2$ and the mass of
bound state $M$, i.e. for any given value of $g^2$ the mass $M$
can be calculated, and vice versa. The results of such calculation
for $\alpha=g^2/4\pi$ are shown in Fig. \ref{pict1} for the case
of scalar meson exchange. The solid curve corresponds to our
calculations in the BS approach, dashed curve represents the
results obtained within Light Front Dynamics \cite{karmanov},
dotted one -- nonrelativistic calculations for the Yukawa
potential.

The spectrum of bound states obtained in this way demonstrates the
customary non-relativistic features. Together with the ground
states, the exited levels of the system can be found (for an alternative
method of calculation see e.g. Ref. \cite{nash_yaf1}). Like in the nonrelativistic picture, solutions of the ground states (i.e. set of
partial vertex functions $g_1,\dots,g_4$) do not have nodes in
$|\bp|$ whereas the excited levels are described by vertex
functions having zeroes.

Besides, for any given mass of exchanged meson $\mu$ the bound
state (at the ground level) can appear in the considered system
only starting from some finite value of the coupling constant
$g_{min}^2$. For example, at $\mu=0.15$ GeV we have
$g^2_{min}=4.023$, which corresponds to some minimal depth of the
potential, where the bound state still exists.

It is obvious, that for weakly bound states for the fixed binding
energies $B$ of order of a few MeV coupling constants are
approximately equal. Thus, for the value $B=1$ MeV
$\alpha_{BS}=0.362$, and $\alpha_{LFD}=0.331$. But for $B$ of
order of hundreds MeV coupling constants essentially differ. For
example, for the value $B=100$ MeV we have the ratio
$\alpha_{BS}/\alpha_{LFD}\sim 1.3$. The explanation of such a
behaviour we found in the role of $^1S_0^{--}$ component in the
total BS solution. It is seen from the Table 1 that its
contribution to the normalization condition is negligibly small
for weakly bound states but increases very rapidly with the
increasing $B$, in contrast to the contributions of other
partial states. Thus, the role of the $^1S_0^{--}$ component is
repulsive, which leads to an increasing coupling constant
for the same bound mass $M$ in comparison with the nonrelativistic
or LFD formalisms, where $''--''$ components are absent.

Another important question to be touched upon is the problem of
stability of the bound states within the BS approach. In the
present context stability means the existence of the solutions in
(\ref{sphom}) without any cut off in the vertices. In its turn,
such a solution exists only if it does not depend on the
parameters of calculation like $N_{max}$, $N_G$, etc. In general,
we found that convergence of our method is quite rapid, and it
becomes faster for smaller values of the coupling constant. In
particular, in the $^1S_0$ channel of the scalar meson exchange kernel it
is sufficient to take into account $N_{max}\sim 4-5$ terms in
(\ref{s0p1})-(\ref{s0p2}) for the values of meson mass $\mu>0.1$
GeV and coupling constant $g^2<40$. In the direct
calculations  it also appears, that the Gaussian 
mesh with $N_G=64$ is almost always
enough, since an increase of $N_G$ beyond this value practically
does not change the results.

Nevertheless, it should be pointed out that at certain conditions
convergence of the solutions becomes poor or even breaks. In
particular, it is lost at small meson masses $\mu \sim 0$, and
similar behaviour was found in \cite{fb-tjon} for the bound states
of scalar particles. In this case, the introduction of form factors allows one
to improve the  situation. However, in total, we found that in general the
equation (\ref{sphom}) has stable solutions only for coupling
constants $g^2$ below some critical value $g_{cr}^2$, which
depends on the type of interaction and the channel considered. It can
be found from the numerical calculations, since at coupling
constants above some critical value the solution disappears, i.e. it
becomes strongly dependent on $N_{max}$, $N_G$ and other
numerical parameters.

\begin{figure}[t]
\vskip -1cm \hskip -0.7cm
\includegraphics[height=8cm]{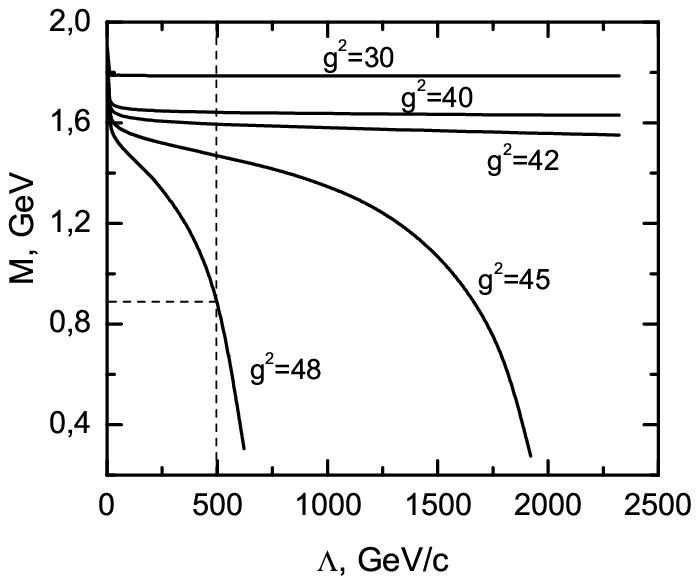}
\hskip -1cm
\includegraphics[height=8cm]{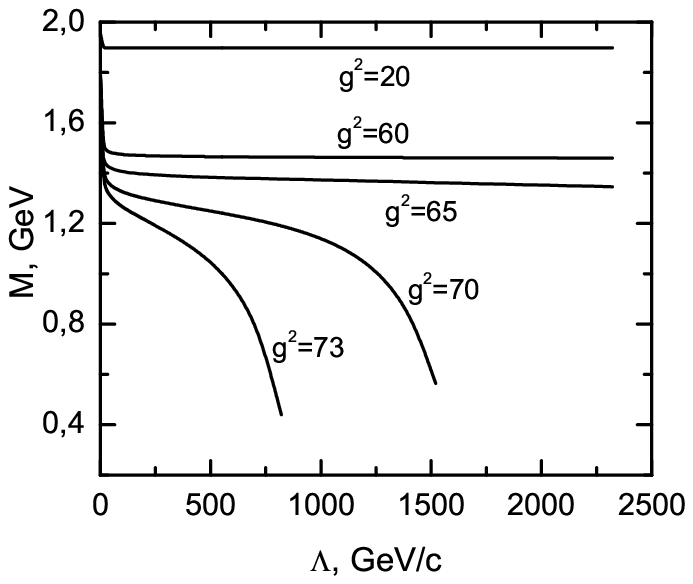}
\\
\parbox[c]{8cm}{\raggedright\protect\caption{Masses of bound
states in $^1S_0$ channel as functions of cut-off parameter
$\Lambda$.}\label{pict3}} \hfill
\parbox[c]{8cm}{\raggedright\protect\caption{Masses of bound
states in $^3S_1$-$^3D_1$ channel as functions of cut-off
parameter $\Lambda$.}\label{pict4}}
\end{figure}

To find the critical value of the coupling constant the dependence of
the ground state mass on the cut-off parameter $\Lambda$ at fixed
$g^2$ have been investigated. The obtained results are presented
in Figs. \ref{pict3} and \ref{pict4} for $^1S_0$ and
$^3S_1$-$^3D_1$ channels respectively. It is evident that in the
limit $\Lambda\to\infty$ and coupling constants $g^2<40$ the
$^1S_0$ bound state mass does not depend on $\Lambda$. The same is
valid for solutions (vertex functions). More exactly, the critical
constant $g^2_{cr}$ for $^1S_0$ state is found to be 40.3.
Theoretical estimation of the critical coupling constant can also
be performed, and it gives the value $g_{cr}^2=4\pi^2\approx 40$.
As it is seen from  Fig. \ref{pict4}, the similar situation
holds for the $^3S_1$-$^3D_1$ channel, where the critical value is
$g_{cr}^2\approx 65$.

It can also be shown that $g^2$ is directly connected to the
power of decrease of vertex functions at large $p$, which is shown
at Fig. \ref{pict2}. The solid line there corresponds to the solution
below $g_{cr}$ with finite normalization, and the dashed one
reproduces the solution beyond critical coupling, which seems to
have infinite normalization.

\section{Conclusion and acknowledgements}
A new method of solving the BS equations for the bound states of
spinor particles by using the expansion of the vertex functions
over the complete set of four-dimensional hyperspherical harmonics
is suggested. Within this method the BS equation is treated in a
ladder approximation for the cases of scalar, pseudoscalar and
vector meson exchanges with corresponding form factors. This
method is shown to be effective and stable.

We are grateful to V.A. Karmanov for valuable discussions
concerning the problem of stability. S.M.D. and S.S.S. acknowledge
the warm hospitality of the Elementary particle physics group of
the University of Rostock, where part of this work was performed.
This work is supported by the Heisenberg - Landau program of JINR
- FRG collaboration and by the Deutscher Akademischer
Austauschdienst.

\end{document}